\documentclass[prd,twocolumn,noshowpacs,preprintnumbers,amsmath,amssymb,nofootinbib]{revtex4}
\usepackage{bm,calrsfs}
\def\a{\alpha}

\def\m{\mu}

\def\be{\begin{equation}}
\def\ee{\end{equation}}
\def\beq{\begin{eqnarray}}
\def\eeq{\end{eqnarray}}

\begin{document}

\title{Algebraic Classification of Weyl Anomalies in Arbitrary Dimensions}

\author{Nicolas Boulanger}
\affiliation{\\ Universit\'e de Mons-Hainaut, Acad\'emie Wallonie-Bruxelles,\\
M\'ecanique et Gravitation, Avenue du Champ de Mars 6, B-7000 Mons, Belgium}

\begin{abstract} 

Conformally invariant massless field systems involving only dimensionless parameters
are known to describe particle physics at very high energy.  
In the presence of an external gravitational field, 
the conformal symmetry may generalize to Weyl invariance. 
However, the latter symmetry no longer survives after quantization: A Weyl
anomaly appears. In this Letter, a purely algebraic understanding of the 
universal structure of the Weyl anomalies is presented. The results hold
in arbitrary dimensions and independently of any regularization scheme. \end{abstract}

\maketitle

At very high energies, like e.g. in 
the early Universe, all the particles can be considered 
as massless, and renormalized matter models are invariant 
under the conformal group. 
Since the Weyl transformations are the generalization to curved space 
of conformal transformations in flat space, 
there are good reasons to anticipate the Weyl symmetry as 
a symmetry of a fundamental theory incorporating gravity~\cite{Fradkin:1985am}. 

On the other hand, symmetries may be broken at the quantum level: 
Anomalies then appear.  
The cancellation of anomalies puts severe constraints on 
the physical content of a theory, as is the case with the 
Standard Model (for a review on anomalies in quantum field theory, see
e.g. \cite{Bertlmann:1996xk}). 
In the case of (super)string theory, the critical dimensions
correspond to the absence of the two-dimensional Weyl anomaly~\cite{Polyakov:1981rd}. 

The Weyl (or conformal, or trace) anomalies have been discovered about 30 years 
ago~\cite{Capper:1974ic,Deser:1976yx} and still occupy a central position in 
theoretical physics, partly because of their important r\^oles within the AdS/CFT 
correspondence and their many applications in cosmology, particle physics, 
higher-dimensional conformal field theory, supergravity and strings.  
The body of work devoted to this subject is, therefore, considerable. 
A very non-exhaustive list of references can be found, e.g., 
in~\cite{Birrell:1982ix,Duff:1993wm,Osborn:1993cr,Bastianelli:2006rx} 

The central equations which determine the candidate anomalies in quantum 
field theory are the Wess-Zumino (WZ) consistency conditions~\cite{Wess:1971yu}. 
By using these conditions, the general structure of all the know anomalies 
\emph{except the Weyl ones} has been determined by purely algebraic methods 
featuring descent equations \`a la Stora-Zumino~\cite{Stora:1976kd,Zumino:1983ew}. 
Such algebraic treatments are crucial since they are 
independent of any regularization scheme and very general. 
The algebraic analysis of anomalies can best be performed within 
the Becchi-Rouet-Stora-Tyutin (BRST~\cite{BRST}) formulation. 

The BRST formulation for the determination of the Weyl anomalies 
was initiated in the pioneering works~\cite{Bonora:1983ff,Bonora:1985cq}, 
with explicit results up to spacetime dimension~$n=6\,$ and the general
structure guessed in arbitrary even dimension.
The authors of~\cite{Bonora:1983ff,Bonora:1985cq} found that the Weyl 
anomalies comprise 
(i) the integral over spacetime of the Weyl scaling parameter times the 
Euler density of the manifold, plus 
(ii) terms that are given by (the integral of) 
the Weyl parameter times strictly Weyl-invariant scalar densities.
Some of the terms from (ii) can trivially be obtained from contractions of 
products of the conformally invariant Weyl tensor,  
while the others are more complicated and involve covariant 
derivatives of the Riemann tensor.  
It was also mentioned in \cite{Bonora:1985cq} that an algebraic
analysis of the Weyl anomalies, similar to the Stora-Zumino treatment 
of the non-Abelian chiral anomaly in Yang-Mills theory, was unlikely to  exist.    

Somewhat later, by using dimensional regularization, 
the authors of~\cite{Deser:1993yx} confirmed the structure of the Weyl anomalies 
found in~\cite{Bonora:1983ff,Bonora:1985cq} and extended the results
to arbitrary (even) dimensions.
The Euler term from class (i) was called ``type\,-A Weyl anomaly'', 
while the terms of (ii) were called ``type\,-B anomalies''.  
Very interestingly, they discovered a similitude between 
the type\,-A Weyl anomaly and the non-Abelian chiral anomaly. 
Accordingly, they hinted at the existence of an algebraic treatment for 
the Weyl anomaly, featuring descent equations. 

In this Letter, we provide for the Weyl anomalies the general, 
purely algebraic understanding \`a la Stora-Zumino that all the other 
known anomalies in quantum field theory enjoy, 
thereby filling a gap in the literature. 
\vspace*{.2cm}

The Weyl anomaly being a local functional, \textit{i.e.} the integral over the 
$n\,$-dimensional spacetime manifold ${\cal{M}}_n$ of a local $n\,$-form $a^n_1$ at ghost
number unity, $gh(a^n_1)=1$ (cf.~\cite{Barnich:2000zw}),  
the WZ consistency conditions for the Weyl 
anomalies~\cite{Bonora:1983ff,Bonora:1985cq} can be written in terms of 
local forms: 
\begin{eqnarray}
&\left\{ 
\begin{array}{cl}
s_{\!_W} a^n_1 + d \,b^{n-1}_{2} =& 0\;,\\
s_{\!_D} a^n_1 + d \,c^{n-1}_{2} =& 0\;,
\end{array} 
\right.\qquad\qquad\qquad\qquad\qquad\qquad&
\label{firsta} \\
&\qquad \qquad \left\{ 
\begin{array}{cl}
a^n_1 \neq &  s_{\!_W} p_0^{n}+d\, f_{1}^{n-1} \\
 \forall \;\;p_0^n & {\rm{s.t.}}\;\; s_{\!_D} p_0^n + d \,h_{1}^{n-1}=0\;.
\end{array}
\right.&
\label{seconda}
\end{eqnarray}
The BRST differentials $s_{\!_W}$ and $s_{\!_D}$ implement the Weyl transformations  
and the diffeomorphisms, respectively, whereas $d$ denotes the exterior total derivative. 
Together with the invertible spacetime metric $g_{\mu\nu}$, the other fields of the problem 
are the Weyl ghost $\omega$ and the diffeomorphisms ghosts $\xi^{\mu}\,$, $\,gh(\xi^{\mu})=gh(\omega)=1\,$. 
The BRST transformations on the fields $\Phi^A=\{g_{\mu\nu},\omega,\xi^{\mu}\}$ read 
\begin{eqnarray}
&s_{\!_D} g_{\mu\nu} = \xi^{\rho}\partial_{\rho}g_{\mu\nu}+\partial_{\mu}\xi^{\rho}g_{\rho\nu}
+\partial_{\nu}\xi^{\rho}g_{\mu\rho}\,,
\;s_{\!_W} g_{\mu\nu} = 2\omega g_{\mu\nu}&
\nonumber \\
&s_{\!_D} \xi^{\mu} = \xi^{\rho}\partial_{\rho}\xi^{\mu}\,,\quad  
s_{\!_D} \omega = \xi^{\rho}\partial_{\rho}\omega\,,
\quad s_{\!_W} \xi^{\mu} = 0 = s_{\!_W} \omega\,.& 
\nonumber 	
\end{eqnarray}    
It should be understood, throughout this Letter, that the space in which BRST-cohomologies 
are to be computed is the space of local $p\,$-forms $b^p$, that is,
the (jet) space of spacetime $p\,$-forms that depend on the fields $\Phi^A$ 
and their derivatives up to some finite (but otherwise unspecified) order, 
which one denotes~\cite{Barnich:2000zw} by 
$b^p=\frac{1}{p!}\,d x^{\mu_1}\ldots d x^{\mu_p}\,b_{\mu_1\ldots\mu_p}(x,[\Phi^A])\,$. 

One unites 
the differentials $s = s_{\!_W}+s_{\!_D}$ and $d$ into a single differential 
$\tilde{s} = s + d\,$, therefore working with local total forms. 
The latter are, by definition, formal sums of local forms with different form degrees 
and ghost numbers, ${\alpha}=\sum_{p=0}^n a^p_{G-p}\,$, the total degree $G$ 
being simply the sum of the form degree and the ghost number.

Powerful techniques for the computation of local BRST cohomologies in top form degree
are exposed in~\cite{Brandt:1996mh} 
and allow one to consider local total forms depending only on a subset ${\cal{W}}$ of the set of 
local total forms, such that $\tilde{s}{\cal{W}}\subset {\cal{W}}$.  
For the general class of theories studied here, the corresponding space ${\cal{W}}$ 
was obtained in~\cite{Boulanger:2004eh}. 

Accordingly, denoting $\tilde{s}_{\!_W}=s_{\!_W}+d$ and similarly for $s_{\!_D}$, 
the problem (\ref{firsta})--(\ref{seconda}) amounts to determining the 
$\tilde{s}_{\!_D}$-invariant $(n+1)$-local total forms $\alpha({\cal{W}})$ satisfying
\begin{eqnarray}
	&\tilde{s}_{\!_W} \alpha({\cal{W}}) = 0 \,, \quad
	\alpha({\cal{W}})\neq \tilde{s}_{\!_W}\zeta({\cal{W}})+constant\,,&
  \label{cohoproblemweyl}
\end{eqnarray}
where $\zeta(\cal{W})$ must be $\tilde{s}_{\!_D}$-invariant.  

Thanks to very general results explained in~\cite{Brandt:1996mh}, 
we know that the solution of (\ref{cohoproblemweyl}) will take the form 
\begin{eqnarray}
\alpha({\cal{W}})=2\omega\,\tilde{C}^{N_1}\ldots\tilde{C}^{N_n}\,a_{N_1\ldots N_{n}}({\cal{T}})\,.&
	\label{ansatz}
\end{eqnarray}
The space ${\cal{T}}$ is generated by~\cite{Boulanger:2004eh} the (invertible) 
metric $g_{\mu\nu}$ together with the $W$-tensors $\{W_{\Omega_i}\}$, 
$i\in\mathbb{N}$, whose precise form will
not be needed here. 
For the purposes of the present Letter, it suffices to know that they contain the
conformally invariant Weyl tensor $W^{\mu}_{~\;\nu\rho\sigma}$
and its first covariant derivative $\nabla_{\tau}W^{\mu}_{~\;\nu\rho\sigma}\,$. 
The symbol $\nabla$ denotes the usual torsion-free metric-compatible covariant 
differential associated with the Christoffel symbols $\Gamma_{~\,\nu\rho}^{\mu}\,$. 
The Ricci tensor is ${\cal{R}}_{\alpha\beta}=R^{\mu}_{~\,\alpha\mu\beta}\,$,  
where $R^{\mu}_{~\;\nu\rho\sigma}=\partial_{\rho}\Gamma_{~\,\nu\sigma}^{\mu}+\ldots$
is the Riemann tensor. 
The scalar curvature is given by ${\cal{R}}=g^{\alpha\beta}{\cal{R}}_{\alpha\beta}\,$. 
Then, one can write the Weyl tensor as
$W^{\mu}_{~\;\nu\rho\sigma} = R^{\mu}_{~\;\nu\rho\sigma}
-2\left(\delta^{\mu}_{\,[\rho}K_{\sigma]\nu}-g_{\nu[\rho}K_{\sigma]}^{~\;\mu} \right)\,$, 
where the tensor 
$K_{\mu\nu} = \frac{1}{n-2}\,\Big({\cal{R}}_{\mu\nu}-\frac{1}{2(n-1)}\,g_{\mu\nu}{\cal{R}} \Big)\,$
plays a key r\^ole in the classification of the Weyl anomalies, as we
will see. Square brackets denote strength-one complete antisymmetrization. 
We also need to recall the definition of the Cotton tensor:  
${C}_{\alpha\rho\sigma} = {2}\,\nabla_{[\sigma}K_{\rho]\alpha}\,$. 

The so-called generalized connections $\tilde{C}^N$ in (\ref{ansatz}) 
are obtained from \cite{Boulanger:2004eh} after setting the 
diffeomorphisms ghosts $\xi^{\mu}$ to zero. 
They read explicitly
\begin{eqnarray}
&\{\tilde{C}^N\} = \{ 2\omega\,,d{x}^{\nu}\,,
 \tilde{C}_{~\,\nu}^{\mu}\,,\tilde{\omega}_{\alpha} \}\,,& 
\nonumber \\
&\tilde{C}_{~\,\nu}^{\mu}=\Gamma_{~\,\nu\rho}^{\mu}\,d{x}^{\rho}\,,\quad
\tilde{\omega}_{\alpha}=\omega_{\alpha} - K_{\alpha\rho}\,d{x}^{\rho}\,, \quad
\omega_{\alpha}=\partial_{\alpha}\omega\,.&
\nonumber 
\end{eqnarray}

As anticipated, the generalized connections $\tilde{\omega}_{\alpha}$ play a crucial 
r\^ole in the classification of the Weyl anomalies. 
They decompose into a ghost part $\omega_{\alpha}$ and a ``connection'' one-form component
${\cal{A}}_{\alpha}=-K_{\alpha\rho}\,dx^{\rho}\,$.  
The decomposition of $\tilde{s}_{\!_W}$ with respect to the 
$\tilde{\omega}_{\alpha}$-degree is at the core of the descent giving the 
type\,-A Weyl anomalies. 
The differential $\tilde{s}_{\!_W}$ decomposes into a part noted 
$\tilde{s}_{\flat}$ which lowers the 
$\tilde{\omega}_{\alpha}$-degree by one unit, a part $\tilde{s}_{\natural}$ 
which does not change the $\tilde{\omega}_{\alpha}$-degree and a part noted 
$\tilde{s}_{\sharp}$ which raises the 
$\tilde{\omega}_{\alpha}$-degree by one unit.

Before displaying the action of $\tilde{s}_{\!_W}$ on ${\cal{W}}$, we need to introduce
some further objects: 
(i) the two-forms $W^{\mu}_{~\;\nu}=\frac{1}{2}\,d x^{\rho}d x^{\sigma}\,W^{\mu}_{~\;\nu\rho\sigma}\,$,
$R^{\mu\nu}=\frac{1}{2}\,dx^{\rho}dx^{\sigma}R^{\mu\nu}_{~~~\rho\sigma}\,$ 
and ${C}_{\alpha}=\frac{1}{2}\,d x^{\rho}d x^{\sigma}\,{C}_{\alpha\rho\sigma}\,$,
(ii) the symbol ${\cal{P}}^{\mu\alpha}_{\rho\,\nu}=(-g^{\mu\alpha}g_{\rho\nu}+
\delta^{\mu}_{\rho}\delta^{\alpha}_{\nu}+\delta^{\mu}_{\nu}\delta^{\alpha}_{\rho})\,$, 
(iii) the generators ${\Delta^{\mu}{}_{\nu}}$ of $GL(n)$-transformations of world indices 
acting on a type\,-$(1,1)$ tensor $T_{\alpha}^{\beta}$ as
${\Delta^{\mu}{}_{\nu}}T_{\alpha}^{\beta}=\delta_{\alpha}^{\mu}T_{\nu}^{\beta}-\delta_{\nu}^{\beta}T_{\alpha}^{\mu}\,$,
and 
(iv) the Weyl-covariant operator ${\cal{D}}_{\mu}=\nabla_{\mu}+K_{\mu\alpha}\mathbf{\Gamma}^{\alpha}$.  
The definition of the generators $\mathbf{\Gamma}^{\alpha}$ is not needed here and 
can be found in~\cite{Boulanger:2004eh}.   
These generators enter the formula for the Weyl transformation of the $W$-tensors:  
$s_{\!_W}W_{\Omega_i}=\omega_{\alpha} \mathbf{\Gamma}^{\alpha}W_{\Omega_i}\,$. 
Both the Cotton two-form ${C}_{\alpha}$ and the generalized connection 
$\tilde{\omega}_{\alpha}$ take their values along the generators $\mathbf{\Gamma}^{\alpha}$,  
${{\mathbf{C}}}={C}_{\alpha}\mathbf{\Gamma}^{\alpha}$ and 
$\tilde{{\omega}}=\omega_{\alpha}\mathbf{\Gamma}^{\alpha}$. 
The Weyl two-form takes its values along the $GL(n)$ generators: 
${\mathbf{W}}=W^{\mu}_{~\;\nu}{\Delta^{\nu}{}_{\mu}}\,$. 
Finally, we denote by $\varepsilon^{\mu_1\ldots\mu_n}$ the totally antisymmetric 
Levi-Civita weight\,-1 density. 
 
Then, the action of $\tilde{s}_{\!_W}$ on ${\cal{W}}$ is given in Table \ref{ta2}, 
following a decomposition with respect to the $\tilde{\omega}_{\alpha}$-degree.  
\begin{table}[h]
\begin{center}
\begin{tabular}{|c||c|c|c|}
\hline
     & $\tilde{s}_{\flat}$ &  $\tilde{s}_{\natural}$  &  $\tilde{s}_{\sharp}$ \\
\hline \hline
$\tilde{\omega}_{\a}$ & \;${C}_{\alpha}$\; & 
$\tilde{C}_{~\,\alpha}^{\beta}\,\tilde{\omega}_{\beta} $ & $0$ \\
\hline
$\omega$ & $0$ & $ 0 $ & $dx^{\mu} \tilde{\omega}_{\mu}$ \\
\hline
$W_{\Omega_i}$ & $ 0 $ & $\tilde{C}_{~\,\nu}^{\mu} {\Delta^{\nu}{}_{\mu}}
W_{\Omega_i}+dx^{\mu}{\cal{D}}_{\mu}W_{\Omega_i}$ & 
$\tilde{\omega}_{\alpha} \mathbf{\Gamma}^{\alpha}W_{\Omega_i}$ \\
\hline
$g_{\alpha\beta}$ & $ 0 $ & $\tilde{C}_{~\,\nu}^{\mu}{\Delta^{\nu}{}_{\mu}}\,g_{\alpha\beta}+2\omega \,g_{\alpha\beta}$ & $0$ \\
\hline
$\tilde{C}_{~\,\nu}^{\mu}$ & 0 & 
$W_{~\;\nu}^{\mu}-\tilde{C}_{~\,\alpha}^{\mu}\tilde{C}_{~\,\nu}^{\alpha}$ & 
  ${\cal{P}}^{\mu\alpha}_{\rho\,\nu}\;\tilde{\omega}_{\alpha}\;d{x}^{\rho}$ \\
\hline
\end{tabular}
\caption{Action of $\tilde{s}_{\!_W}$, decomposed w.r.t the $\tilde{\omega}_{\a}$-degree \label{ta2}}
\end{center}
\end{table}

We can now state the following two theorems, the central results reported in this Letter: 
\vspace*{.1cm}

{\underline{\it{Theorem 1\,:}}} ~
Let $\psi_{\m_1\ldots\m_{2p}}$ be the local total form 
\begin{eqnarray}
	\psi_{\mu_1\ldots\mu_{2p}} &=& 
	\frac{\omega}{\sqrt{-g}} \;
	\varepsilon^{\alpha_1\ldots\alpha_r}_{\quad\quad~\,\nu_1\ldots\nu_r\mu_1\ldots\mu_{2p}}
	\nonumber \\
	&& \qquad\qquad \times ~ \tilde{\omega}_{\alpha_1}\ldots\tilde{\omega}_{\alpha_r}
	\;d x^{\nu_1}\ldots d x^{\nu_r}\,,
	\nonumber \\
	p&=&m-r\,,\quad m=n/2\,,\quad 0\leqslant r\leqslant m\,
  \nonumber
\end{eqnarray}
and $W^{\mu\nu}$ the tensor-valued two-form
	$W^{\mu\nu} = W^{\mu}_{~\;\rho}\,g^{\rho\nu}\,$.
Then, the local total forms ${\Phi}_r^{[n-r]}$ $(0 \leqslant r \leqslant m)$
\begin{eqnarray}
	\Phi^{[n-r]}_{r} = \frac{(-1)^p}{2^p}\,\frac{m!}{r!\,p!}\;\psi_{\m_1\ldots\m_{2p}}\,
	W^{\m_1\m_2}\ldots \,W^{\m_{2p-\!1}\m_{2p}}
\nonumber
\end{eqnarray}
obey the descent of equations
\begin{eqnarray}
&\left\{
\begin{array}{cl}
	\tilde{s}_{\flat}\Phi^{[n-r]}_{r} + \tilde{s}_{\natural}\Phi^{[n-r+1]}_{r-1} &=\; 0\quad,
	\nonumber \\
	\tilde{s}_{\sharp}\Phi^{[n-r]}_{r} &=\; 0\quad,\quad (1\leqslant r\leqslant m)
	\nonumber 
\end{array}\right.&
\\
  &\tilde{s}_{\flat}\Phi^{[n-1]}_{1} 
  \;=\; 0 \;=\; \tilde{s}_{\!_W} \Phi^{[n]}_{0}\;,&	
\nonumber
\end{eqnarray}
so that the following relations hold: 
	$\tilde{s}_{\!_W} {\alpha} = 0 = \tilde{s}_{\!_W}{\beta} \,$,
	with  
	${\alpha}=\sum_{r=1}^{m}\Phi^{[n-r]}_{r}\,$ and ${\beta} = \Phi^{[n]}_{0}\,$. 
\vspace*{.3cm}

{\underline{\it{Theorem 2\,:}}} ~
(A) The top form-degree component $a^n_1$ of ${\alpha}$ (cf. Theorem 1) 
satisfies the WZ consistency conditions for the Weyl anomalies. 
The WZ conditions for $a^n_1$ give rise to a non-trivial descent 
and $a^n_1$ is the \textit{unique} anomaly with such a property,  
up to the addition of trivial terms and anomalies satisfying a trivial descent. 

(B) The top form-degree component $e^n_1$ of $(\alpha+\beta)$
is proportional to the Euler density of the manifold ${\cal{M}}_{n}\,$: 
\begin{eqnarray}
{e}^n_1 &=& \frac{(-1)^m}{2^m}\; 
\sqrt{-g}\;\omega \,(R^{\mu_1 \nu_1}\ldots\, R^{\mu_m \nu_m})\;
\varepsilon_{\mu_1 \nu_1\ldots \,\mu_m \nu_m}\;.
\nonumber  	
\end{eqnarray}
Clearly, the anomaly $\beta=\Phi^{[n]}_{0}$ satisfies a trivial descent 
since it is given by a contraction of a product of Weyl tensors 
($m$ of them in dimension $n=2m$).
\vspace*{.3cm}

{\underline{\it{Proofs\,:}}} ~
The existence part of the non-trivial descent problem for the Weyl anomalies 
is given in Theorem 1 and part (B) of Theorem 2. 
It is proved by direct computation. 
Only part (A) of Theorem 2, the uniqueness part of the problem, is not straightforward. 
The detailed proof is given elsewhere~\cite{Boulanger:2007st}. 
It follows lines of reasonings as in e.g.~\cite{Barnich:1995ap,Barnich:2000zw,Brandt:1996au} 
and uses general results given in~\cite{Barkallil:2002fp}.  
The essential point is to determine the most general expression at the bottom of 
the non-trivial descents associated with the Weyl anomalies. 
(The anomalies that satisfy the trivial descent $s_{\!_W} a^{n}_{1}=0$ 
are the type\,-B Weyl anomalies~\cite{Deser:1993yx};  
they can be classified along the lines of~\cite{Boulanger:2004zf,Boulanger:2004eh}.)
It turns out~\cite{Boulanger:2007st} 
that the most general element at the bottom of these descents is the component
of $\Phi^{[m]}_m$ with maximal ghost number $m+1\,$.
\vspace*{.2cm}

We now illustrate our two theorems 
with the descents corresponding to $n=2$, $4$ and $6\,$. 
The general case can readily be understood from these three examples. 
\vspace*{.2cm}

The case $n=2$ is a bit special. Although $K_{\mu\nu}$ is not determined
($\sim\frac{0}{0}$), its trace $K_{\rho}^{~\rho}={\cal{R}}/(2n-2)$ is well-defined.  
Theorem 1 gives $\Phi^{[2]}_0 = 0$ and
$\alpha=\Phi^{[1]}_1 = \frac{\omega}{\sqrt{-g}}\,\varepsilon^{\mu\rho}g_{\rho\nu}\tilde{\omega}_{\mu}dx^{\nu}$
$=\omega\sqrt{-g}\,\varepsilon_{\rho\nu}g^{\rho\mu}\tilde{\omega}_{\mu}dx^{\nu}$. 
Taking the top form degree of $\alpha\,$, 
we find
$a^2_1=\frac{\omega}{2}\,\sqrt{-g}\,{\cal{R}}\,d^{2}x$, the well-known result for the Weyl
anomaly in two dimensions. 
\vspace*{.1cm}

Next, using Theorem 1 in the case $n=4$ gives 
\begin{eqnarray}
\Phi^{[4]}_0 &=& \frac{\omega}{4}\,\sqrt{-g}\;\varepsilon_{\mu_1\ldots\mu_4}\;
W^{\mu_1\mu_2}W^{\mu_3\mu_4}\,,
\nonumber \\
\Phi^{[3]}_1 &=& -\,\omega\,\sqrt{-g}\;\varepsilon^{\alpha}_{~\,\nu\rho\sigma}\;
\tilde{\omega}_{\alpha}\,dx^{\nu}\,W^{\rho\sigma}\,,
\nonumber \\
\Phi^{[2]}_2 &=& \omega\,\sqrt{-g}\;\varepsilon^{\alpha\beta}_{~\,~\,\rho\sigma}\;
\tilde{\omega}_{\alpha}\tilde{\omega}_{\beta}\,dx^{\rho}dx^{\sigma}\,. 
\nonumber 
\end{eqnarray}
The top form-degree component of $(\alpha + \beta)$ is $e^4_1\,$: 
\begin{eqnarray}
e^4_1 &=& 
\frac{\omega}{4}\,\sqrt{-g}\;\varepsilon_{\mu\nu\rho\sigma}
(W^{\mu\nu}-2\,{\cal{A}}^{\mu}dx^{\nu})(W^{\rho\sigma}-2\,{\cal{A}}^{\rho}dx^{\sigma})
\nonumber
\end{eqnarray}
which obviously reproduces the expression for the Euler term of Theorem 2
because of the following identities:   
\begin{equation}
R^{\mu\nu} = W^{\mu\nu}-2\,{\cal{A}}^{[\mu}dx^{\nu]}\;,
\quad {\cal{A}}^{\mu} = - g^{\mu\nu}K_{\nu\rho}dx^{\rho}\;.
\label{imporela}
\end{equation}
The descent for $n=4$ thus reads
\begin{eqnarray}
&& \left\{  
\begin{array}{cl}
	s_{\!_W}e^4_1 + d \, b^3_2 &= \;0 \quad,
\nonumber \\
  s_{\!_W}b^3_2 + d \, b^2_3 &= \;0 \quad,
\nonumber \\
  s_{\!_W}b^2_3 &= \;0 \quad,\qquad {\rm{with}}
\nonumber 
\end{array}\right.
\\
b^{3}_2 &=& -2\,\omega\,\sqrt{-g}\;\varepsilon^{\alpha}_{~\,\nu\rho\sigma}\;
{\omega}_{\alpha}\,K_{~\,\mu}^{\nu}\,dx^{\mu}dx^{\rho}dx^{\sigma}\,,
\nonumber \\
b^{2}_3 &=& \omega\,\sqrt{-g}\;\varepsilon^{\alpha\beta}_{~\,~\,\rho\sigma}\;
{\omega}_{\alpha}{\omega}_{\beta}\,dx^{\rho}dx^{\sigma}\,. 
\nonumber 
\end{eqnarray}

Finally, in dimension $6$, Theorem~1 and Theorem~2 give (a representative of)
the unique Weyl anomaly satisfying a non-trivial descent of equations: 
\begin{eqnarray}
e^6_1 &=& 
\frac{-\omega}{8}\,\sqrt{-g}\;\varepsilon_{\mu_1\ldots\mu_6}
R^{\mu_1\mu_2}R^{\mu_3\mu_4}R^{\mu_5\mu_6}\,.
\label{e6}
\end{eqnarray}
The elements of the corresponding descent are obtained, as before, 
\textit{via} the $\Phi_r^{[n-r]}$'s of Theorem 1:  
\begin{eqnarray}
\beta=\Phi^{[6]}_0 &=& \frac{-\omega}{8}\,\sqrt{-g}\;\varepsilon_{\mu_1\ldots\mu_6}\;
W^{\mu_1\mu_2}W^{\mu_3\mu_4}W^{\mu_5\mu_6}\,,
\nonumber \\
\Phi^{[5]}_1 &=& \frac{3\,\omega}{4}\,\sqrt{-g}\;
\varepsilon^{\alpha}_{~\,\nu\mu_1\ldots\mu_4}\;
\tilde{\omega}_{\alpha}\,dx^{\nu}\,W^{\mu_1\mu_2}W^{\mu_3\mu_4}\,,
\nonumber \\
\Phi^{[4]}_2 &=& \frac{-3\,\omega}{2}\,\sqrt{-g}\;
\varepsilon^{\alpha\beta}_{~\,~\,\mu\nu\rho\sigma}\;
\tilde{\omega}_{\alpha}\tilde{\omega}_{\beta}\,dx^{\mu}dx^{\nu}
\,W^{\rho\sigma}\,,
\nonumber \\
\Phi^{[3]}_3 &=& \omega\,\sqrt{-g}\;
\varepsilon^{\alpha\beta\gamma}_{\quad~\,\mu\nu\rho}\;
\tilde{\omega}_{\alpha}\tilde{\omega}_{\beta}\tilde{\omega}_{\gamma}\,
dx^{\mu}dx^{\nu}dx^{\rho}\;. 
\nonumber
\end{eqnarray}
Extracting from $\alpha=\Phi^{[5]}_{1}+\Phi^{[4]}_{2}+\Phi^{[3]}_{3}$
its top form-degree component amounts to
selecting everywhere the contribution ${\cal{A}}_{\mu}$ of 
$\tilde{\omega}_{\mu}=\omega_{\mu}+{\cal{A}}_{\mu}$.  
As a consequence, the top form-degree component of $(\alpha+\beta)$ 
reproduces the expression (\ref{e6}), 
making use of the identities (\ref{imporela}). 
On the other hand, extracting the different ghost-number components
of $\alpha$
provides us with the elements $b^5_2$, $b^4_3$ and $b^3_4$ of the descent for $e^6_1$:  
\begin{eqnarray}
&& \left\{  
\begin{array}{cl}
	s_{\!_W}e^4_1 + d \, b^5_2 &= \;0 \quad,
\nonumber \\
  s_{\!_W}b^5_2 + d \, b^4_3 &= \;0 \quad,
\nonumber \\
  s_{\!_W}b^4_3 + d \, b^3_4 &= \;0 \quad,
\nonumber \\
  s_{\!_W}b^3_4 &= \;0 \quad. 
\nonumber 
\end{array}\right.
\end{eqnarray}
Without the addition of the type\,-B anomaly $\beta$, the top form-degree 
component $a_1^{6}$ of $\alpha$, taken alone, gives
\begin{eqnarray}
	a_1^{6} &=& \frac{-3\,\omega}{8}\,\sqrt{-g}\;\varepsilon_{\mu_1\ldots\,\mu_6}
	 \Big[  (-2\,{\cal{A}}^{\mu_1}dx^{\mu_2})W^{\mu_3\mu_4}W^{\mu_5\mu_6}
 \nonumber \\
	 && +\; (-2\,{\cal{A}}^{\mu_1}dx^{\mu_2})(-2\,{\cal{A}}^{\mu_3}dx^{\mu_4})W^{\mu_5\mu_6}
 \nonumber \\
	 && +\;
	 (-2\,{\cal{A}}^{\mu_1}dx^{\mu_2})(-2\,{\cal{A}}^{\mu_3}dx^{\mu_4})
	 (-2\,{\cal{A}}^{\mu_5}dx^{\mu_6}) \Big ]\;.
	 \nonumber
\end{eqnarray}

As we have shown, adding $\beta=\Phi^{[n]}_0$ to $a^{n}_1$ somehow
``covariantizes'' the latter, producing the Euler term $e^n_1\,$. 
The Weyl anomaly $a_1^{n}$ 
is reminiscent of the consistent non-Abelian chiral anomaly. 
However, note that the descent for $a_1^{n}$ stops at form-degree 
$\frac{n}{2}\,>0\,$. 
Amusingly, the Euler form $e^n_1$ looks like the non-Abelian singlet anomaly.
The ``trace over the internal indices'' is taken with the Levi-Civita density.  

{\underline{\it{Conclusions$\,$:}}}~ 
The universal structure
of the Weyl anomalies is established in a purely algebraic manner, 
independently of any regularization scheme and in arbitrary
dimensions. In particular, we do not resort to dimensional analysis. 
The type$\,$-A Weyl anomaly of~\cite{Deser:1993yx} is 
the counterpart of the consistent non-Abelian chiral anomaly, 
in that it is the \textit{unique} Weyl anomaly 
satisfying a non-trivial descent of equations. 
This solves a long-standing problem and answers a question originally 
due to Deser and Schwimmer~\cite{Deser:1993yx}. 
Since the Weyl anomalies associated with a trivial descent 
can be systematically built and classified as in~\cite{Boulanger:2004zf,Boulanger:2004eh}, 
our analysis completes a general, purely algebraic classification 
of the Weyl anomalies in arbitrary spacetime dimensions. 

This work was supported by the Fonds de la Recherche Scientifique, FNRS (Belgium). 
We thank G. Barnich for many useful discussions and M. Henneaux for having
suggested the project. We thank H. Osborn and Ph. Spindel for their comments and 
encouragements.

\end{document}